\documentclass[aps,twocolumn,amsmath,showpacs]{revtex4}
  \usepackage[dvips]{color}
  \usepackage{newlfont}
  \usepackage{graphicx}
  \usepackage{amssymb}
  \usepackage{amsmath}
  \usepackage[latin1]{inputenc}
  \usepackage{float}

\begin{document}

\title{Semiclassical limit of the entanglement in closed pure 
systems}

\author{Renato M. Angelo}
\email{renato@ifi.unicamp.br}
\affiliation{ Instituto de Física, Universidade de São Paulo,\\
C.P. 66318, 05315-970, São Paulo, SP, Brazil.}
\author{K. Furuya}
\email{furuya@ifi.unicamp.br}
\affiliation{Instituto de Física `Gleb Wataghin', Universidade
Estadual de Campinas, \\ C.P. 6165, 13083-970, Campinas, SP, 
Brazil}

\date{\today}

\begin{abstract}
We discuss the semiclassical limit of the entanglement for the 
class of closed pure systems. By means of analytical and 
numerical calculations we obtain two main results: (i) the 
short-time entanglement does not depend on Planck's constant and
(ii) the long-time entanglement increases as more semiclassical
regimes are attained. On one hand, this result is in contrast with 
the idea that the entanglement should be destroyed when the macroscopic 
limit is reached. On the other hand, it emphasizes the role 
played by decoherence in the process of emergence of the classical 
world. We also found that, for Gaussian initial states, the 
entanglement dynamics may be described by an entirely classical 
entropy in the semiclassical limit.
 

\end{abstract}

\maketitle
\section{Introduction}\label{sec:introd}
 
Understanding the emergence of all aspects of the classical world 
from the quantum theory is far from being a trivial task. In fact, 
it is a recurrent question which has been asked since the early days of 
quantum mechanics. However, at the level of the expectation values of 
position and momentum, the classical limit is reasonably understood 
\cite{berman,berry,ballentine,angelo}, according to the conceptual 
scheme we try to sketch in the next few lines. 

Concerning the Newtonian trajectories, 
which are based on the assumption of complete knowledge of the initial 
conditions, their emergence from the quantum expectation values is 
predicted by the Ehrenfest theorem, valid for initially localized states. 
Within the Ehrenfest time scale, expectation values agree with the 
classical trajectories and, in this sense, the quantum mechanics 
recovers the determinism of the Newtonian theory. The classical limit ($
\hbar\to 0$) in this case is in general asymptotic and does not require 
any further mechanism such as decoherence. In other words, there exists
a formal classical limit for expectation values even in the context of 
closed systems. Although such a mathematical limit is not unique, one 
possible way to obtain it is particularly convenient: 
a two-degree of freedom classical Hamiltonian may be produced with 
the assumption of a separable dynamics of coherent states \cite{saraceno}.
In this approach, the classical theory emerges as a consequence of
the absence of the entanglement.

After the Ehrenfest time scale, the Newtonian mechanics is no longer 
able to mimic the quantum expectation values of observables and
the single trajectory determinism is missed. 
It means that predictions can only be made by means of averages over 
ensembles. But, even in this case, classical equations of motion are 
applicable (within the Liouville formalism). This is ensured by a
mechanism known as {\sl decoherence} \cite{decoherence}. Environment, to 
which all systems are inevitably coupled, destroys quantum correlations 
and turns quantum predictions identical to those offered by the classical 
statistical formalism. Since the decoherence time is in general much 
smaller than the Ehrenfest time, the correspondence between quantum and 
classical world will be complete for all instants of the dynamics.

Within the above described lines of thought, a natural step before 
the introduction of the action of the environment is to analyze the 
classical limit of an intrinsically quantum quantity - the {\sl entanglement}
 - for the class of {\sl closed pure systems}, for which the 
theoretical semiclassical limit ($ \hbar\to 0$) is well established at the 
level of expectation values. 
Here, a word concerning the entanglement is in order, recalling that the 
entanglement is the greatest resource offered ``exclusively'' by the 
quantum world, which allows for the realization of the most challenging 
ideas, like quantum information processing and quantum computation 
\cite{bennett,vedral, nielsen}. Since entanglement is regarded as an 
intrinsically quantum property with no classical analog, it seems to be an 
adequate modern tool for the study of the semiclassical behavior of quantum 
mechanics.

We then ask about the behavior of the entanglement in such situations in 
which an underlying classical dynamics is well-known. This question motivates 
us to analyze the semiclassical limit of the entanglement dynamics 
for closed pure systems in which classicality is achieved by means of 
macroscopic coherent states, i.e., minimal uncertainty states $|\alpha\rangle$ 
with $|\alpha|²\gg 1$. 

 At a first sight, one may be induced to expect a decrease in the 
entanglement in the semiclassical limit.
However, we show analytically and also by means of numerical examples that 
this is not a precise idea: in fact, the short-time entanglement is 
$\hbar$-independent for closed systems (section II).

This paper is organized as follows: in section II we discuss a short time 
expansion for the entanglement in an arbitrary pure bipartite system. In 
Section III a semiclassical analysis of the whole dynamics of the entanglement
 is numerically investigated for two different models and an explanation 
in terms of mechanisms that are classically understandable as well is given. 
Section IV is reserved for some concluding remarks. 
\section{The Short-Time Entanglement}\label{sec:analytical}

For globally pure bipartite systems there exist suitable entropic quantities 
obeying all requirements for an adequate entanglement measure 
\cite{vedral2,manfredi}. Particularly useful is the reduced linear entropy, 
which offers in general the same informations given by the von Neumann 
entropy but with a much less computational effort.

The reduced linear entropy (RLE) for a bipartite system
belonging to the Hilbert space $\cal{E}=\cal{E}_u\otimes
\cal{E}_v$ is defined by 
\begin{eqnarray}
S(t)=1-\textrm{Tr}_{z}[\rho_{z}²(t)],
\label{S}
\end{eqnarray}
in which $\rho_{u,v}(t)=\textrm{Tr}_{v,u}[\rho(t)]$ is the reduced
density operator. The index $z$ denotes the subsystems ``$u$'' and ``$v$''. 
The density operator $\rho$ satisfies the von Neumann equation 
$\imath\hbar \dot{\rho}=[H,\rho]$, being $H$ the Hamiltonian 
operator. This is a dynamical measure of the purity of the 
subsystem $z$ in the unitary bipartite dynamics of initially pure 
states. For this case, the Schmidt decomposition \cite{schmidt} 
guarantees that RLE is also a measure of the entanglement 
between the parts \cite{angelo2}. Furthermore, the Araki-Lieb 
inequality \cite{araki} guarantees that the RLE of the subsystems
 are identical for pure states.

In order to obtain a semiclassical expansion for the entanglement 
in the approximation of short times we write the Dyson's series 
for the density operator up to the order $t²$:
\begin{eqnarray}
\rho(t)\simeq\rho_0+\frac{[H,\rho_0]}{\imath \hbar}\,t+\frac{[H,[H,
\rho_0]]}{2\,(\imath\hbar)²}\,t²,\label{Dyson}
\end{eqnarray}
where $\rho_0=|u_0\rangle\langle u_0|\otimes|v_0\rangle\langle v_0|$ 
is an initially disentangled pure state composed by the product of 
bosonic coherent states corresponding to the subsystems $u$ and $v$. 
Now we apply the partial trace over the subsystem $u$ in the coherent 
state basis to get
\begin{subequations}
\begin{eqnarray}
\rho_v(t)\simeq \rho_v^{(0)}+\rho_v^{(1)}\,t+\frac{1}{2}\,\rho_v^{(2)}\,t²,
\end{eqnarray}
\begin{eqnarray}
\rho_v^{(0)}&=&|v_0\rangle\langle v_0|, \\
\rho_v^{(1)}&=&\int\frac{d²u_1}{\pi}\frac{\langle u_1|[H,\rho_0]
|u_1\rangle}{\imath\hbar}, \\
\rho_v^{(2)}&=&\int\frac{d²u_1}{\pi}\frac{\langle u_1|[H,[H,\rho_0]]|u_1
\rangle}{(\imath\hbar)²}.
\end{eqnarray}
\label{rhov}
\end{subequations}
Then, the RLE defined by \eqref{S} may be written as
\begin{subequations}
\begin{eqnarray}
S(t)\simeq S^{(0)} + S^{(1)} t +S^{(2)} t²,\label{St2}
\end{eqnarray}
\begin{eqnarray}
S^{(0)}&=&1-\textrm{Tr}_v\left[\rho_v^{(0)}\right],\\
S^{(1)}&=&-\textrm{Tr}_v\left[\rho_v^{(0)}\rho_v^{(1)}+\rho_v^{(1)}\rho_v^
{(0)} \right],\\
S^{(2)}&=&-\textrm{Tr}_v\left[\frac{\rho_v^{(0)}\rho_v^{(2)}+\rho_v^{(2)}
\rho_v^{(0)}}{2}+\rho_v^{(1)}\rho_v^{(1)}  \right],
\end{eqnarray}
\end{subequations}
where $\textrm{Tr}_v[\,\cdot\,]=\int \frac{d²v_1}{\pi}\langle v_1|\cdot|
v_1\rangle$. 

Straightforward manipulations on this equations yield $S^{(0)}=S^{(1)}=0$. 
Consequently, $S(t)\simeq S^{(2)}t²$. This result, which has already 
appeared in literature \cite{kim}, is a direct consequence of both the 
purity and the separability of the initial state.

Defining the dimensionless Hamiltonian
\begin{eqnarray}
\mathbb{H}\equiv \frac{H\,t}{\hbar},
\label{H}
\end{eqnarray}
we put \eqref{St2} in the following compact form:
\begin{eqnarray}
S(t) \simeq 2\,\Big[\cal{C}_{00}(\mathbb{H})+\cal{C}_{11}
(\mathbb{H})-\cal{C}_{10}(\mathbb{H})-\cal{C}_{01}(\mathbb{H})  
\Big],
\label{SH}
\end{eqnarray}
in which he have defined the correlations
\begin{subequations}
\begin{eqnarray}
\cal{C}_{00}&=&\Big(\langle\psi_0|\,\mathbb{H}\,|\psi_0\rangle 
\Big)², \\
\cal{C}_{01}&=&\langle \psi_0|\,\mathbb{H}\,\rho_{1}(0)\,
\mathbb{H}\,|\psi_0\rangle, \\ \cal{C}_{10}&=&\langle \psi_0|\,
\mathbb{H}\,\rho_{2}(0)\,\mathbb{H}\,|\psi_0\rangle, \\ 
\cal{C}_{11}&=&\langle \psi_0|\,\mathbb{H}²\,|\psi_0\rangle,
\end{eqnarray} \label{C}
\end{subequations}
with $|\psi_0\rangle=|u_0\rangle\otimes|v_0\rangle$.
Interestingly, by \eqref{SH} we may put the conditions for the 
existence of short-time entanglement in terms of the inequality
\begin{eqnarray}
 2\,\Big(\cal{C}_{00}+\cal{C}_{11}-\cal{C}_{10}-\cal{C}_{01}  
\Big)>0.
\label{ineq}
\end{eqnarray}

 Now we look for a semiclassical expansion for the short-time 
entanglement given by \eqref{SH} in the basis of coherent states. 
We start by defining a general two-degree of freedom classical 
Hamiltonian.
\begin{eqnarray}
\cal{H}(q_u,p_u,q_v,p_v)=\sum\limits_{n,m,l,k}c_{nmlk}q_u^n\,
p_u^m\,q_v^l\,p_v^k. \label{Hc}
\end{eqnarray}
This is an explicit $\hbar$-independent function.
Following the ordered quantization procedure \cite{angelo} 
we obtain the corresponding Hamiltonian operator
\begin{eqnarray}
H=\cal{S}_{u}\,\cal{S}_{v}\,\sum\limits_{n,m,l,k}c_{nmlk}Q_u^n\,
P_u^m\,Q_v^l\,P_v^k, \label{Hq}
\end{eqnarray}
with $[Q_z,P_z]=\imath \hbar$. $\cal{S}_{z}=\exp{\left(-\frac{\imath
\hbar}{2}
\partial_{Q_z}\partial_{P_z} \right)}$ is the symmetric ordering 
operator \cite{angelo}. This is a suitable quantization
method for our purposes since it factorizes the dependences in 
$\hbar$. There are other sources of $\hbar$, namely, the operator
$\mathbb{H}$ and the parameterizations of the coherent states for the
phase space. 

The calculations of all terms in \eqref{SH} are made by inserting
\eqref{Hq} in \eqref{C}. Then, using the unities of the coherent 
states basis, namely, $\mathbf{1}_u=\int \frac{d²u_1}{\pi}|u_1\rangle
\langle u_1|$ and $\mathbf{1}_v=\int \frac{d²v_1}{\pi}|v_1\rangle
\langle v_1|$,
we are lead to 
\begin{eqnarray}
\frac{\hbar^2}{2 t^2}S(t)&=&\Big(\langle u_0 v_0|H|u_0 v_0\rangle\Big)^2 
\nonumber\\&+&\int\frac{d^2u_1}{\pi}\frac{d^2 v_1}{\pi}\langle u_0 v_0|H
|u_1 v_1\rangle \langle u_1 v_1|H|u_0 v_0\rangle  \nonumber \\
&-&\int\frac{d^2u_1}{\pi}\langle u_0 v_0|H|u_1 v_0\rangle \langle u_1 v_0
|H|u_0 v_0\rangle  \nonumber \\&-&\int\frac{d^2v_1}{\pi}{\pi}\langle u_0 
v_0|H|u_0 v_1\rangle \langle u_0 v_1|H|u_0 v_0\rangle,\label{<S>}
\end{eqnarray}
The function $\langle u_0 v_0|H|u_1 v_1\rangle$ is the most general kernel 
we have to manipulate in the calculation of the correlations. By
 \eqref{Hq} we obtain
\begin{eqnarray}
\langle u_0 v_0|H|u_1 v_1\rangle&=&\sum\limits_{n,m,k,l}c_{nmkl}\langle u_0
| \cal{S}_u Q_u^n P_u^m|u_1\rangle \nonumber \\ & & \times\langle v_0|
\cal{S}_v Q_v^k P_v^l|v_1\rangle.\label{<H>}
\end{eqnarray}
The properties of the symmetric ordering operator \cite{angelo} 
allow us to write
\begin{subequations}
\begin{eqnarray}
\frac{\langle u_0| \cal{S}_u Q_u^n P_u^m|u_1\rangle}{\langle u_0|u_1\rangle}
= e^{\frac{1}{2}\partial_{u_0^*}\partial_{u_1}}\left[\left(q_u^{(01)} \right
)^n\left(p_u^{(01)} \right)^m \right],\,\,\,\,\,\,\label{<QP>}
\end{eqnarray}
\begin{eqnarray}
q_u^{(ij)}\equiv\frac{\langle u_i|Q_u|u_j\rangle}{\langle u_i|u_j\rangle}, 
\qquad
p_u^{(ij)}\equiv\frac{\langle u_i|P_u|u_j\rangle}{\langle u_i|u_j\rangle},
\end{eqnarray}
\end{subequations}
with similar expressions for $v$. Notice that $q^{ij}$ and $p^{ij}$
correspond to the usual canonical phase space pair when $i=j$. Otherwise, 
when $i\neq j$, they are complex numbers composed by combinations of 
distinct canonical pairs.

Given the usual parametrization of the coherent state label $u$ (or $v$) 
for the phase space (Eq.\eqref{v}) we may write $\partial_u=\sqrt{\frac{
\hbar}{2}}\left(\partial_{q_u}-\imath\partial_{p_u} \right)$. Considering 
smooth wave packets at $t=0$, we may regard this partial derivative as a 
small parameter and the exponential operator in \eqref{<QP>} may be 
expanded in few terms. After performing such adequate expansions in 
\eqref{<QP>} we return to \eqref{<H>} to calculate the semiclassical 
expansion of the kernel.

The next step is to calculate the integrals in \eqref{<S>}, which is done 
by means of the following useful formula \cite{angelo}:
\begin{eqnarray}
\int\frac{d^2 u}{\pi}|\langle u_0|u\rangle|^2 f(u_0,u)=\left[e^{\frac{\hbar}
{4}\nabla_u^2} f(u_0,u)\right]_{u=u_0},
\end{eqnarray}
with $\nabla^2_u=\partial_{q_u}^2+\partial_{p_u}^2$. This relation emerges 
from the fact that $|\langle u_0|u\rangle|^2$ is a Gaussian weight function. 
Once again we may get a semiclassical expression by expanding the exponential 
operator in the first orders in $\hbar$. In fact, the procedure described up 
to here must be performed in  a consistent way such that the short-time RLE 
may be written as
\begin{eqnarray}
S(t)\simeq t² \Big[\cal{O}(\hbar^{-2})+\cal{O}(\hbar^{-1})+\cal{O}(\hbar^0)
+\cdots \Big],\label{SO}
\end{eqnarray}
Notice that the classical limit $\hbar\to 0$ seems to yield a divergence 
of the short-time entanglement, what would be a quite unexpected result. 
However, an exhaustive calculation shows that $\cal{O}(\hbar^{-2})=\cal{O}
(\hbar^{-1})=0$ and $\cal{O}(\hbar^0)\neq 0$, in general. This derivation 
is really lengthy and tedious, but does not require any further recipe 
beyond that one presented above and it will be omitted here. 

Concluding this calculation, in a semiclassical regime, where $\hbar$ is 
finite, but arbitrarily smaller than a typical action of the system, the 
RLE is given by
\begin{eqnarray}
S(t)\,\simeq\, t²\, \cal{O}(\hbar^0),
\label{Sth}
\end{eqnarray}
showing that the short-time entanglement {\sl does not depend} on $\hbar$ 
in a semiclassical regime.
 
The constant $\cal{O}(\hbar^0)$ is constituted by a sum of terms involving 
first-order derivatives (in phase space variables) of the Hamiltonian 
function \eqref{Hc}. Note that, the possibility of expressing the first 
contributions to the entanglement dynamics in terms of classical quantities 
is a direct consequence of the special basis used in the calculation.

Result \eqref{Sth}, which is general within the class of bipartite bosonic 
systems in pure states, attests that the entanglement does not tend 
asymptotically to the classically expected limit $S(t)=0$, as does those
well-behaved quantities like expectation values of canonical operators $Q$ 
and $P$.

Also it has to be remarked that the quadratic dependence in time found for 
the short-time entanglement is independent of the integrability of the 
underlying classical system, i.e., it is always algebraic no matter whether 
the classical system is chaotic or regular. In an approach based 
on the Loschmidt echo dynamics \cite{thomas} it has been predicted different
behavior for the entanglement in chaotic and regular regimes. However,
this has not been done for the time scale we are considering, 
and furthermore, in that case the entanglement generation is 
obtained by means of two distinct unitary evolutions (echo operator).
In this sense, there is no conflict with our results. 
In fact, the quadratic dependence of the short-time linear entropy 
has already been found in \cite{kim}. It is just a consequence 
of the initial separability of the quantum state. Recent studies 
also confirm our results concerning the algebraic dependence 
\cite{jacquod} and the $\hbar$-independence \cite{prosen} of the 
short-time entanglement.

Next we present two numerical examples confirming the above 
result and also discuss the long-time behavior of the entanglement.
\section{Numerical Analysis}

\subsection{Classical States}

Since we are interested in the quantum-to-classical aspect of the
entanglement, we avoid initial states that are initially entangled, 
mixed or delocalized. The natural choice is the minimum uncertainty 
coherent states defined by 
\begin{subequations}
\begin{eqnarray}
|v\rangle=D(v)\,|r\rangle,
\end{eqnarray}
\begin{eqnarray}
D(v)= \left\{ \begin{array}{ll}
\displaystyle{ \exp{[v\hat{a}^+-v^{*}\hat{a}]}} ,\\ \\
\displaystyle{ \exp{\left[\frac{\arctan|v|}{|v|}(v\hat{J_+}-v^{*}
\hat{J_-})\right]}}.
\end{array}\right. \label{D}
\end{eqnarray} \label{|v>}
\end{subequations}
$D$ was defined respectively for the harmonic oscillator and for 
the angular momentum (spins). The reference state $|r\rangle$ 
stands for the ground states $|0\rangle$, in the Fock basis, and 
$|J,-J\rangle$, in the angular momentum basis.

The coherent state label, $v$, has the following common 
parametrization in terms of classical phase space variables:
\begin{eqnarray}
v= \left\{ \begin{array}{ll}
\displaystyle{\frac{q+\imath\,p}{\sqrt{2\hbar}} } ,\\ \\
\displaystyle{\frac{q+\imath\,p}{\sqrt{\hbar\,J-(q²+p²)}}}.
\end{array}\right.\label{v}
\end{eqnarray}

For bosonic coherent states, the classicality is attained by 
taking $|v|²$ sufficiently macroscopic to guarantee that the 
average energy $\langle v|H|v\rangle$ be much greater than a 
typical spectral distance $E_{n+1}-E_{n}$. In this case, taking 
$|v|²\to\infty$ is mathematically equivalent to requiring 
$\hbar\to 0$. On the other hand, for spin coherent states, the 
classicality emerges by means of a more sophisticated limit  
\cite{SLDicke}: $J\to\infty$, $\hbar\to 0$ and $\hbar
\sqrt{J(J+1)}\approx \hbar J =1$.

We consider Hamiltonian operators like 
\begin{eqnarray}
H=H_1\otimes\mathbf{1}_2+\mathbf{1}_1\otimes H_2+H_{12},
\label{Ham}
\end{eqnarray}
the last term being a nonlinear coupling which is capable to 
entangle initially separable coherent states.

\subsection{Dicke Model}

First, we investigate the entanglement in one of the most important 
models in quantum optics (QO) and more recently in quantum computation (QC), 
namely, the $N$-atom Jaynes-Cummings Model (or Tavis-Cummings model) 
\cite{tavis}. 
In QO it is mostly known in the so called rotating wave approximation (RWA), 
and describes the coupling of a monochromatic field of frequency $\omega$ 
with $N$ non-interacting two-level atoms with level separation $\epsilon$.
Recently in connection with QC, the model received renewed attention from 
solid-state community working on `phonon cavity quantum dynamics' 
\cite{weig04,vorrath04}, Josephson Junctions and quantum dots \cite{lee04}, 
quantum chaos \cite{emary} and quantum phase transitions \cite{lambert}. 
These works are interested on the model in the original form conceived by 
Dicke \cite{dicke} without the RWA. The Hamiltonian in the last case is 
composed by the following terms
\begin{eqnarray}
\begin{array}{l}
\displaystyle{H_1=\epsilon\,J_z,\qquad H_2=\hbar\,\omega\,
a^{\dag}a,} \\ \\
\displaystyle{H_{12}=\frac{G}{\sqrt{2J}}\left(aJ_++a^{\dag}J_- 
\right)+\frac{G'}{\sqrt{2J}}\left(a^{\dag}J_++aJ_- \right),}
\end{array}\label{HNJCM} 
\end{eqnarray}
in which the operators $a$ and $a^{\dag}$ are the usual harmonic 
oscillator creation and destruction operators associated to the field, 
and $\hat{J_+}, \hat{J_-},\hat{J_z}$ are angular momentum ladder 
operators. Here, the spin algebra is associated to the atoms, being the 
total spin given by $J=N/2$. $G$ and $G'$ are real coupling constants,
and generally could be taken unequal.

This model possesses a classical counterpart 
$\cal{H}_{\textrm{\tiny $N$-JCM}}(J)$ obtained
in ref. \cite{aguiar} which presents several interesting properties, 
e.g., chaotic behavior. 
Another peculiar feature is the scaling property
\begin{eqnarray}
\frac{\cal{H}_{\textrm{\tiny $N$-JCM}}(J)}{J}=\frac{\cal{H}_{
\textrm{\tiny $N$-JCM}}(J')}{J'}.\label{scaling}
\end{eqnarray}
If this relation is satisfied for the pairs $[E(J),J]$ and 
$[E(J'),J']$, then the associated dynamics will be totally 
equivalent. 

The initial state $|\psi_0\rangle=|v_a\rangle\otimes|v_f\rangle$ 
 is thus constructed by means of the following process. We use the 
formula $\mathbf{r}_J=\mathbf{r}_{1}\,\sqrt{J}$ in order to 
re-scale the vector of initial conditions $\mathbf{r}_J=(q_a,p_a,
q_f,p_f)$, keeping the dynamics for $J=1$ as a reference. This
 guarantees automatically that we obtain a dynamics $[E(J),J]$ 
equivalent to the unitary $[E(1),1]$. Then, we construct the 
coherent initial state by putting $\mathbf{r}_J$ in \eqref{v} 
and \eqref{|v>}. Resuming, we chose initial quantum states with 
different $J$, but producing always the same classical dynamics. 
The semiclassical conditions in this case are satisfied by 
increasing $J$ and keeping $\hbar$ constant (as it indeed occurs in nature). 
It may be shown that this is totally equivalent to the mathematical 
condition $\hbar=1/J$, with $J\to\infty$ and fixed initial conditions.

\begin{figure}[ht]
\centerline{\includegraphics[scale=0.35,angle=-90]{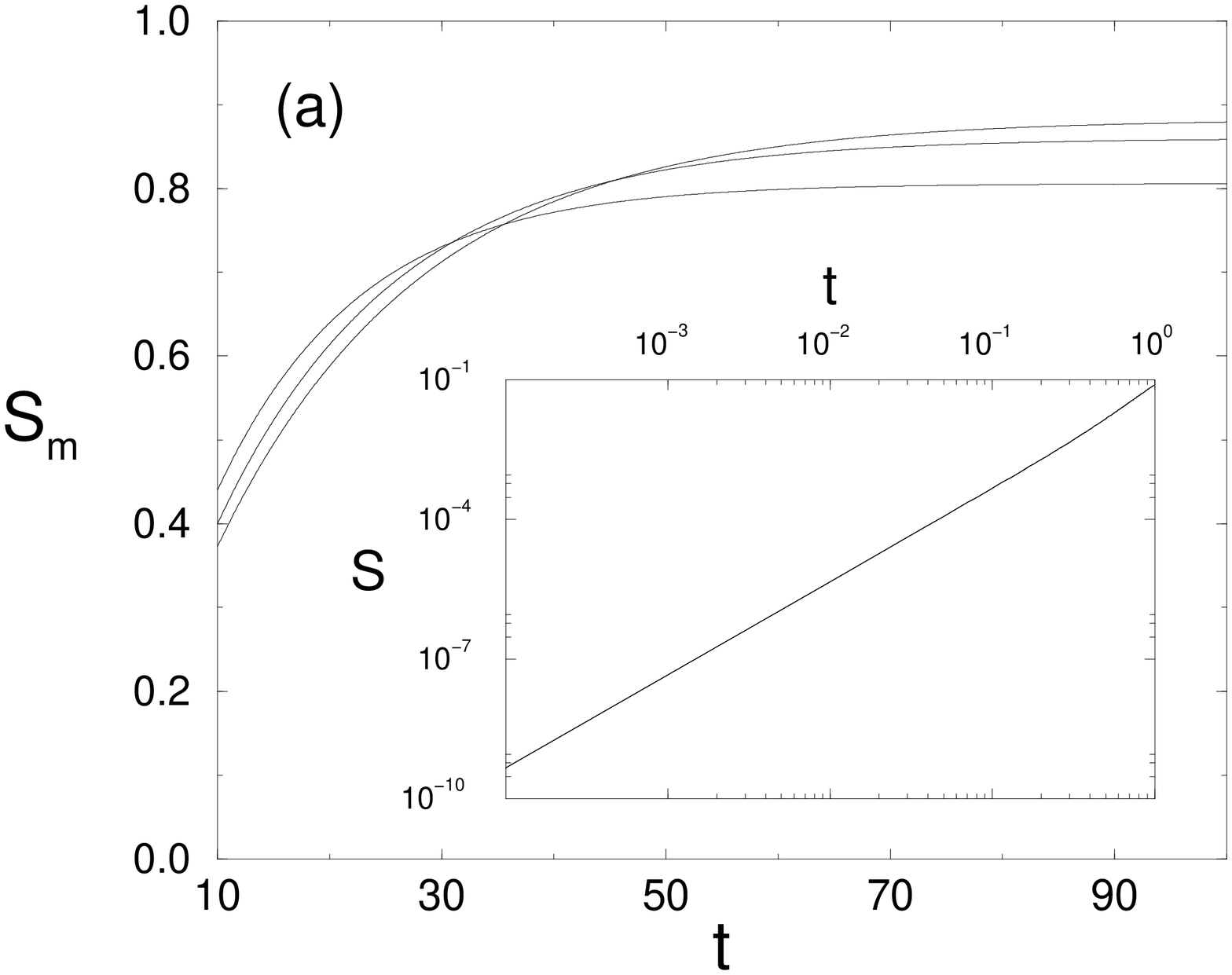}}
\centerline{\includegraphics[scale=0.35,angle=-90]{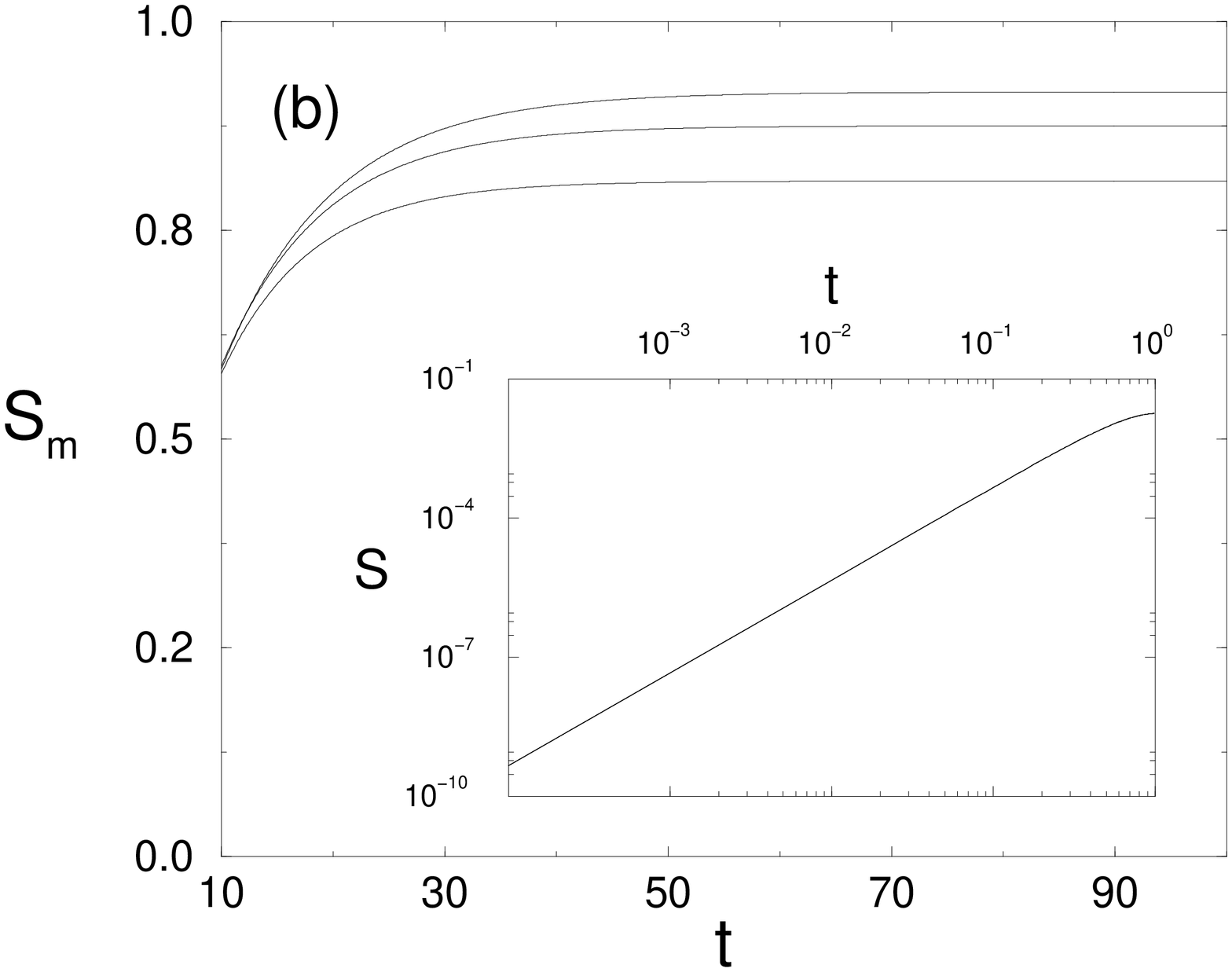}}
\caption{\small Mean long-time entanglement for initial coherent states 
centered in (a) a periodic orbit and in (b) a chaotic trajectory. The insets 
show the result for the short-time entanglement, which behaves like
 $S=0.05\,t²$, for all values of $J$.
Upper curves correspond to larger values of $J$, which assumes the values 
3.5, 6.5 and 10.5. In these calculations we used $\epsilon=\omega= 
1.0$, $G=G'=0.35$ and $\hbar=1$, in arbitrary units.}
\label{SJCM}
\end{figure}

In Fig.\ref{SJCM} we show numerical results for the 
entanglement as a function of time for several values of $J$. 
Actually, for long times, the quantity plotted is a kind 
of mean entanglement 
of the states `associated' to a given classical trajectory. We
calculated it as follows: given a certain classical 
trajectory, we choose a set of $M$ initial states 
$|\psi_i(0)\rangle$ centered at points along the trajectory, 
and calculate the respective RLE's, $S_i(t)$. 
Then, the averaged quantity is given by
\begin{eqnarray}
S_m(t)\equiv\frac{1}{M}\sum\limits_{i=1}^{M}S_i(t).
\end{eqnarray}
Such a calculation allows us to observe a smooth mean 
behavior for entanglement without the characteristic 
oscillations associated to the border effect \cite{angelo3}.
The mean long-time entanglement is then suitably described by a fitting 
expression \cite{angeloD} given by
\begin{eqnarray}
S_m(t)=A_0\left(1-e^{-A_1\,t} \right),
\end{eqnarray}
where $A_0$ and $A_1$ are fitting parameters.

At least two aspects are remarkable in this numerical result: 
(i) the short-time behavior, proportional to $t²$, is 
$J$-independent and (ii) the plateau value increases with $J$.

The aspect (i) has already been predicted in section 
\ref{sec:analytical}, but for a different class of Hamiltonians. 
The results shown in Fig.\ref{SJCM} are, therefore, a numerical 
verification of \eqref{Sth} in a more general context, in 
which the Hamiltonian operator $H$ is not obtained from a classical 
function by means of a quantization based on bosonic coherent 
(Gaussian) states. 
Such a $\hbar$-independence induces us to believe that the rising 
in the entanglement is determined just by classical sources, as 
the local departure of neighboring phase space points. 
This is indeed suggested by previous studies on quantum chaos 
\cite{furuya}. But the main point is that entanglement does 
not diminish as more macroscopic regimes are attained. 

The second aspect may seem even more curious at first sight, since 
it means that the entanglement increases as we tend to a more 
semiclassical limit. We will re-take this analysis after our 
second example.

\subsection{Coupled Nonlinear Oscillators}

We thus proceed to show analytical results of the entanglement 
dynamics in a second model, for which it is possible to understand 
the behavior of the entanglement in the presence of an effect which  
has no classical analog, namely, the quantum interference. This is done 
for a model of two resonant oscillators coupled nonlinearly 
which in certain regimes describe a coupled Bose-Einstein condensates
(BEC). The terms of the Hamiltonian (\ref{Ham}) are
\begin{eqnarray} \begin{array}{l}
H_k = \hbar\,\omega\,\left(a_k^{\dag}\,a_k+\frac{1}{2}\right) 
\qquad (k=1,2),\\
H_{12}=\hbar\lambda \left(\hat{a}_1^{\dag}\hat{a}_2+\hat{a}_1
\hat{a}_2^{\dag}\right)+\hbar^2 g \left(\hat{a}_1^{\dagger}{
\hat{a}_1}+
                   \hat{a}_2^{\dagger}{\hat{a}_2}+1\right)^2.
\end{array}\label{HBEC}
\end{eqnarray}
Such Hamiltonian has been applied to describe an ideal two-mode 
BEC in the equal scattering length situation \cite{leggett} with 
internal Josephson coupling \cite{cirac}. Recently, the entanglement 
dynamics between the modes for initially disentangled coherent states, 
was exactly calculated \cite{lsanz}. We reproduce here the 
analytical result for the RLE:
\begin{eqnarray}
S(t)&=& 1 -e^{ -2|\beta_{1}\left(t\right)|^2}
\sum\limits_{n,m}
\frac{|\beta_{1}\left(t\right)|^{2n}}{n!}\frac{|\beta_{1}
\left(t\right)|^{2m}}{m!} \nonumber \\
&\times& e^{-4|\beta_{2}\left(t\right)|^2\sin^2{\left[\hbar\,g\,t
\left(n-m\right)\right]}},
\label{SBEC}
\end{eqnarray}
in which $ \beta_{1,2}(t)=(\alpha_{1,2}e^{-\imath\omega t}\cos{
\lambda t}- \imath\,\alpha_{2,1}e^{-\imath\omega t}\sin{\lambda 
t})$ and $\alpha_i=(q_i+\imath\, p_i)/\sqrt{2\hbar}$. Result 
\eqref{SBEC} predicts that the system recovers totally its 
purity (disentangled states) at
\begin{eqnarray}
t=\frac{\pi}{g\,\hbar},\label{t}
\end{eqnarray}
after passing by an intricate interference process (see 
\cite{lsanz} for details).

Following the reasoning presented in Sec.\ref{sec:introd}, our 
argumentation will be supported by a classical entropy, 
which is defined within the classical theory of ensembles. It 
describes the classical correlation dynamics of initially 
separable probabilities. This quantity has been shown to 
evolve very closely to its quantum counterpart, the reduced 
linear entropy, in cases that are initially separable and in 
which intrinsic quantum effects are absent \cite{angelo2}. 
Similar approaches may be found in \cite{gong}. 
We then introduce the {\em classical reduced linear entropy} (CRLE), 
defined in Ref.\cite{angelo2} as follows
\begin{eqnarray}
S_{cl}(t)=1-\frac{\text{tr}_k\,[P_k(t)]}{\text{tr}_k\,[P_k(0)]},
\label{Scl}
\end{eqnarray}
in which the partial classical trace is given by $\text{tr}_k=
\int dq_k dp_k$, for $k=1,2$. $P_{1,2}$ denotes the reduced 
probability function obtained by $\text{tr}_{2,1}[P(t)]$, in 
which the joint probability $P(t)$ is a function satisfying the 
Liouville equation
\begin{eqnarray}
\frac{\partial P}{\partial t}=\left\{\cal{H},P\right\}.
\label{Liouville}
\end{eqnarray}
$\cal{H}$ is the classical Hamiltonian function. The 
normalization in \eqref{Scl} is necessary to guarantee both an 
adequate dimensional unity and $S_{cl}(0)=0$. The connection 
between quantum and classical worlds at $t=0$ is established by taking
\begin{eqnarray}
P(0)=\frac{\langle v_1 v_2|\,\rho_0\,|v_2 v_1\rangle}{2\pi\hbar},
\label{P0}
\end{eqnarray}
where $\rho_0$ is the pure density operator at $t=0$. The 
normalization is chosen in such a way that $\text{tr}P=1$. 
The initial distribution \eqref{P0} is the unique source of $\hbar$ 
in the CRLE dynamics. Thus, the semiclassical regime in the classical 
Liouvillian formalism is associated to strongly localized 
initial distributions.

\begin{figure}[ht]
\centerline{\includegraphics[scale=0.35,angle=-90]{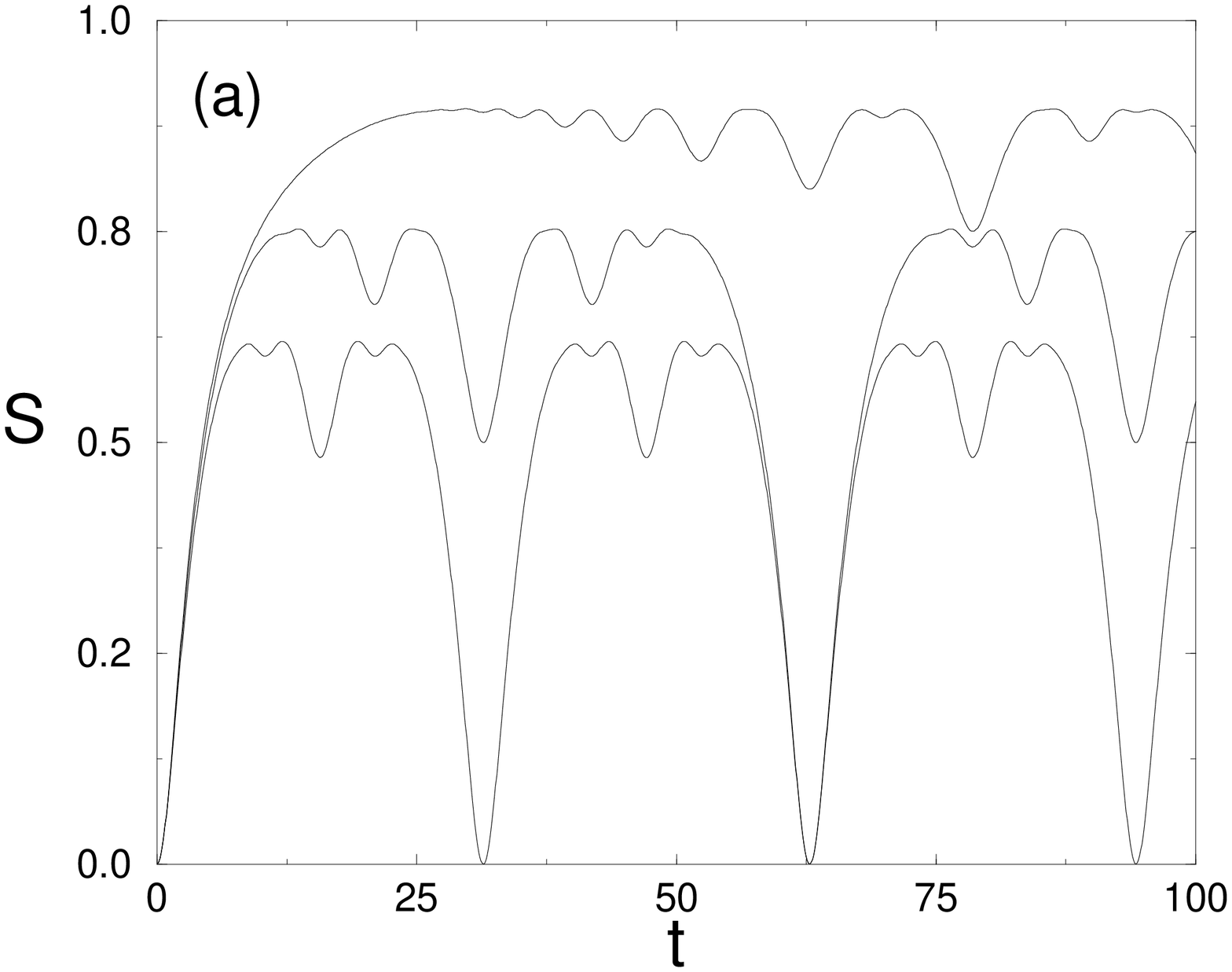}} 
\centerline{\includegraphics[scale=0.35,angle=-90]{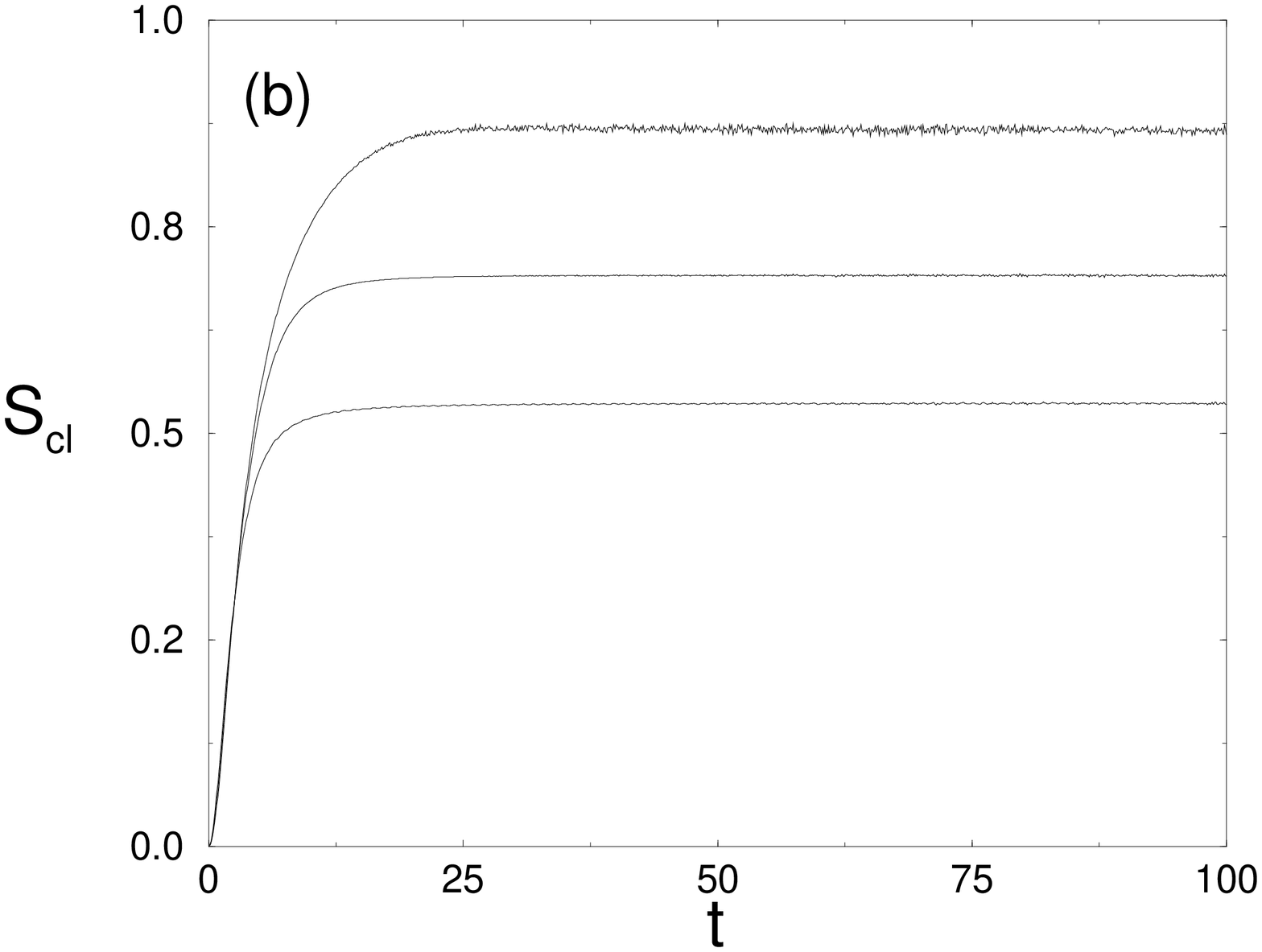}}
\caption{\small (a) Entanglement between two Bose-Einstein 
condensates and (b) the classical reduced linear entropy for the 
 classical counterpart as a function of time. 
Higher plateaus were attained for smaller values of $\hbar$. The 
parameter values used were $\omega=1$, $g=0.1$, $\lambda=0.2$, $q_i=p_i=1$ 
and $\hbar$ assumed the values 0.1, 0.5 and 1.0 (arbitrary units).}
\label{SQCbec}
\end{figure}

The numerical results for both quantum and classical reduced 
linear entropies are shown in Fig.\ref{SQCbec}. The dependence of
CRLE in time reflects the fact that classical probabilities also become 
correlated, i.e., initially independent probability 
distributions are transformed into conditional ones as the 
dynamical evolution takes place. In some cases, similarities 
between these quantum and classical quantities indicate that 
entanglement has indeed a strong statistical character 
\cite{angelo2}. Here, however, remarkable differences emphasize 
the role played by an intrinsic quantum effect, as is the 
interference. In fact, it has been shown that interference is 
the major responsible for such flagrant differences, since 
there is no analog effect in the classical formalism 
\cite{lsanz}.

Concerning the semiclassical limit, we see that quantum 
recurrences, and consequently, quantum interferences, are 
postponed by the dynamics as $\hbar$ is made small as
compared with the classical action $|\beta_1|^2+
|\beta_2|^2$. This is indeed predicted analytically by \eqref{t}. 
Accordingly, the resemblance between quantum and classical entropies 
tends to increase in semiclassical regimes. In this sense, these 
results allows us to consider the CRLE
as the very semiclassical limit of the entanglement. This is an 
indication that in a semiclassical regime, the occurrence of the 
entanglement is due solely to spreading effects of the wave function, 
i.e., statistical effects present also in the classical theory of ensembles.

The entanglement dynamics of the system \eqref{HBEC} presents 
the same qualitative general features of our precedent example, 
namely: short-time behavior independent of $\hbar$ (see \cite{oliveira} 
for the analytical demonstration) and 
higher plateaus for more classical regimes. Actually, this scenario has 
been shown to be quite general for a wide class of nonlinear systems 
\cite{MT}.

These results lead us to separate the dynamics of the entanglement  
in two main time scales. The first one, namely the short-time scale,
may be regarded as a classical scale in which entanglement is 
determined essentially by the vicinity of the initial condition 
in classical phase space. Accordingly, it has been shown that 
this scale is connected with the Ehrenfest time for the system 
\eqref{HBEC} \cite{lsanz}.
Then, this is a situation in which, although 
the subsystems are entangled, the quantum state still allows 
the correspondence between quantum expectation values and classical 
trajectories.

The second scale, associated to long times (for the which the  
Ehrenfest theorem does not apply), 
contains all the allowed quantum effects, like 
interferences, and also indicates the fact that the entropy is 
extensive with the number of pure states accessible to the dynamics. 
These remarkably different time scales and the extensivity of the 
entropy have already been mentioned in slightly different contexts 
in \cite{lsanz} and \cite{angelo4} respectively. Also, other time 
scales have been established in \cite{oliveira}.

All results shown here point out to the same important 
conclusion: in closed pure bipartite systems, entanglement 
exists even in arbitrarily semiclassical regimes. 
On the other hand, in a strictly formal classical limit 
($\hbar=0$), entanglement is indeed expected to disappear 
completely, since classical points have no statistics associated. 
In fact, this is a crucial assumption for obtaining the
Newtonian trajectories from the quantum formalism \cite{saraceno}.
But this is a peculiar mathematical limit which will never be 
accessible in physical world.  In this sense, our results
also indicate that the formal classical limit of entanglement is
rather singular for closed systems.

The analysis presented above concerns a theoretical framework 
for the classical limit of quantum mechanics in closed pure systems. 
It is a mathematical limit which ensures that quantum mechanics is 
a universal theory which recovers the classical results in a particular 
regime of parameters. In closed systems, the limit is asymptotic 
for expectation values but seems to be rather singular for the 
entanglement. However, this conceptual scenario does not match the 
real world of quantum open systems. In this sense, our results must 
be regarded as the starting point for a more fundamental analysis that 
takes into account the effects of the environment, to which every real 
system is coupled. As it is well known, decoherence tends to wash out 
quantum superpositions and the entanglement is expected to disappear 
under such situation. However, a rigorous test of such assertions 
requires an adequate entanglement measure for mixed states of continuous 
variable systems, which is still a very controversial issue.

\section{Concluding remarks}

We have shown by two numerical examples and by analytical 
calculations that entanglement does not decrease as the 
semiclassical regime is attained asymptotically in closed pure 
systems. In fact, the results pointing out to an increasing amount 
of entanglement as the equilibrium is attained.

This behavior may be understood by noticing that semiclassical 
regime avoids quantum interferences, but it is not able to 
eliminate the spreading of the wave packet. Even extremely 
localized coherent states perceive the local spreading 
caused by the classical flux of trajectories associated. 
Furthermore, since we note that the entropy is extensive with the 
number of pure quantum states of the density operator, the increase 
in the long-time entanglement in the semiclassical limit becomes 
just a natural fact.

Recently, it was shown that entanglement can always arise in the 
interaction of an arbitrarily large system in any mixed state 
with a single qubit in a pure state \cite{bose01}. This result 
stresses the role played by the subsystem purity as an enforcer 
of entanglement, even in a thermodynamic limit of high 
temperatures. Here, we are concerned with a different kind of 
classicality, namely, that one attained by $\hbar \to 0$. 
Furthermore, we have studied strictly the case where the subsystems 
are in initially in pure states. 
The common point in these investigations is that the regarded 
systems do not interact with the environment. In this sense, 
{\em unitarity} seems to be a key word in such apparent contradicting 
semiclassical limits. Then, there is a remaining question 
to be answered: Is indeed decoherence able to provide the 
``expected'' semiclassical limit for the entanglement?

\acknowledgments

It is a pleasure to thank G. Q. Pellegrino and M. C. Nemes 
for helpful discussions, remarks and suggestions. We also acknowledge 
Conselho Nacional de Desenvolvimento Científico e Tecnológico (CNPq) 
and Fundação de Amparo à Pesquisa do Estado de São Paulo (FAPESP)
(Contract 02/10442-6) for financial support.


\end{document}